\documentclass[sigconf]{acmart}
\theoremstyle{definition}
\newtheorem{definition}{Definition}

\usepackage{arydshln}
\usepackage{threeparttable}
\usepackage{multirow}

\makeatletter
\renewcommand{\email}[2][]{%
  \g@addto@macro\addresses{}% do nothing
}
\makeatother

\AtBeginDocument{%
  }

\usepackage{nameref}

\newenvironment{ResearchQuestion}[1]
    {
    \vspace{3pt}
    {\narrower \noindent \textit{\textbf{Research Question:} #1}\par}
    \vspace{3pt}
    }

\acmConference[ICSE '26]{2026 IEEE/ACM 48th International Conference on Software Engineering}{April 12--18, 2026}{Rio de Janeiro, Brazil}
\acmBooktitle{2026 IEEE/ACM 48th International Conference on Software Engineering (ICSE '26), April 12--18, 2026, Rio de Janeiro, Brazil}
\begin{document}
\sloppy
\title[The Software Infrastructure Attitude Scale (SIAS)]{The Software Infrastructure Attitude Scale (SIAS): A~Questionnaire Instrument for Measuring Professionals' Attitudes Toward Technical and Sociotechnical Infrastructure}

%%Alternative titles
%\title{Measuring Developers’ Attitudes: Development and Validation of the Software Infrastructure Attitude Scale (SIAS)}


%\author{Miikka Kuutila and Paul Ralph}
%\affiliation{
%  \institution{Dalhousie University}
%  \city{Halifax}
%  \state{Nova Scotia}
%  \country{Canada}
%}
%\email{miikka.kuutila@dal.ca; paulralph@dal.ca}

\author{Miikka Kuutila}
\email{miikka.kuutila@helsinki.fi}
\affiliation{%
  \institution{Dalhousie University}
  \city{Halifax}
  \country{Canada}
}
\affiliation{%
\institution{University of Helsinki,}
\country{Finland}
{\ttfamily miikka.kuutila@\discretionary{}{}{}helsinki.fi}
}

\author{Paul Ralph}
\email{paulralph@dal.ca}
\affiliation{%
  \institution{Dalhousie University}
  \city{Halifax}
  \country{Canada}
  \texttt{paulralph@\,dal.ca}
}

\author{Huilian Sophie Qiu}
\email{sophie.qiu@kellogg.northwestern.edu}
\affiliation{%
  \institution{Northwestern University}
  \city{Evanston}
  \country{USA}
  \texttt{huilian\_qiu@}\\
  \texttt{\,alumni.brown.edu}\\
}

\author{\mbox{Ronnie de Souza Santos}}
\email{ronnie.desouzasantos@ucalgary.ca}
\affiliation{%
  \institution{University of Calgary}
  \city{Calgary}
  \country{Canada}
  \texttt{ronnie.desouzasantos}
  \texttt{@ucalgary.ca}
}

\author{Morakot Choetkiertikul}
\email{morakot.cho@mahidol.ac.th}
\affiliation{%
  \institution{Mahidol University}
  \city{Nakhon Pathom}
  \country{Thailand}
  \texttt{morakot.cho@\,mahidol.ac.th}
}

\author{Amin Milani Fard}
\email{amilanif@nyit.edu}
\affiliation{%
  \institution{New York Institute of Technology}
  \city{Vancouver}
  \country{Canada}
  \texttt{amilanif@\,nyit.edu}
}

\author{Rana Alkadhi}
\email{ralkadi@KSU.EDU.SA}
\affiliation{%
  \institution{King Saud University}
  \city{Riyadh}
  \country{Saudi Arabia}
  \texttt{ralkadi@\,ksu.edu.sa}
}

\author{Xavier Devroey}
\email{xavier.devroey@unamur.be}
\orcid{0000-0002-0831-7606}
\affiliation{%
  \institution{NADI, University of Namur}
  \city{Namur}
  \country{Belgium}
  \texttt{xavier.devroey@\discretionary{}{}{}unamur.be}
}

\author{Gregorio Robles}
\email{grex@gsyc.urjc.es}
\affiliation{%
  \institution{U. Rey Juan Carlos}
  \city{Madrid}
  \country{Spain}
  \texttt{grex@\,gsyc.urjc.es}
}

\author{Hideaki Hata}
\email{hata@shinshu-u.ac.jp}
\affiliation{%
  \institution{Shinshu University}
  \city{Nagano}
  \country{Japan}
  \texttt{hata@\,shinshu-u.ac.jp}
}

\author{Sebastian Baltes}
\email{sebastian.baltes@uni-heidelberg.de}
\affiliation{%
  \institution{Heidelberg University}
  \city{Heidelberg}
  \country{Germany}
  \texttt{sebastian.baltes@}
  \texttt{\,uni-heidelberg.de}
}

\author{Vladimir Kovalenko}
\email{vladimir.kovalenko@jetbrains.com}
\affiliation{%
  \institution{JetBrains, Amsterdam}
 % \city{Amsterdam}
  \country{The Netherlands}
  \texttt{vladimir.kovalenko@}
  \texttt{\,jetbrains.com}
}

\author{Shalini Chakraborty}
\email{shalini.chakraborty@uni-bayreuth.de}
\affiliation{%
  \institution{University of Bayreuth}
  \city{Bayreuth}
  \country{Germany}
  \texttt{shalini.chakraborty}
  \texttt{@uni-bayreuth.de}
}

\author{Eray Tuzun}
\email{eraytuzun@cs.bilkent.edu.tr}
\affiliation{%
  \institution{Bilkent University}
  \city{Ankara}
  \country{Turkey}
  \texttt{eraytuzun@}
  \texttt{\,cs.bilkent.edu.tr}
}

\author{Hera Arif}
\email{hera.arif@dal.ca}
\affiliation{%
  \institution{Dalhousie University}
  \city{Halifax}
  \country{Canada}
  \texttt{hera.arif@\,dal.ca}
}

\author{Gianisa Adisaputri}
\email{gianisa@dal.ca}
\affiliation{%
  \institution{Dalhousie University}
  \city{Halifax}
  \country{Canada}
  \texttt{gianisa@\,dal.ca}
}

\author{Kelly Garc\'{e}s}
\email{kj.garces971@uniandes.edu.co}
\affiliation{%
  \institution{Universidad de los Andes}
  \city{Bogotá}
  \country{Colombia}
  \texttt{kj.garces971@}
  \texttt{uniandes.edu.co}
}

\author{Anielle S. L. Andrade}
\email{anielle.lisboa@gmail.com}
\affiliation{%
  \institution{PUCRS}
  \city{Porto Alegre}
  \country{Brazil}
  \texttt{anielle.lisboa@\,gmail.com}
}

\author{Eyram Amedzor}
\email{ey460215@dal.ca}
\affiliation{%
  \institution{Dalhousie University}
  \city{Halifax}
  \country{Canada}
  \texttt{eyramm@\,dal.ca}
}

\author{Bimpe Ayoola}
\email{Bimpe.Ayoola@dal.ca}
\affiliation{%
  \institution{Dalhousie University}
  \city{Halifax}
  \country{Canada}
  \texttt{Bimpe.Ayoola@\,dal.ca}
}

\author{\mbox{Keisha Gaspard-Chickoree}}
\email{Keisha.Gaspard-Chickoree@dal.ca}
\affiliation{%
  \institution{Dalhousie University}
  \city{Halifax}
  \country{Canada}
  \texttt{Keisha.Gaspard}
  \texttt{-Chickoree@\,dal.ca}
}

\author{Arazoo Hoseyni}
\email{a.hoseyni@dal.ca}
\affiliation{%
  \institution{Dalhousie University}
  \city{Halifax}
  \country{Canada}
  \texttt{a.hoseyni@\,dal.ca}
}

\renewcommand{\shortauthors}{Kuutila et al.}

\begin{abstract}
\textbf{Context:} Recent software engineering (SE) research has highlighted the need for sociotechnical research, implying a demand for customized psychometric scales.
\textbf{Objective:} We define the concepts of technical and sociotechnical infrastructure in software engineering, and develop and validate a psychometric scale that measures attitudes toward them.
\textbf{Method:} Grounded in theories of infrastructure, attitudes, and prior work on psychometric measurement, we defined the target constructs and generated scale items. The items were reviewed and refined by domain experts. The scale was administered to 225 software professionals and evaluated using a split sample. We conducted an exploratory factor analysis (EFA) on one half of the sample to uncover the underlying factor structure and performed a confirmatory factor analysis (CFA) on the other half to validate the structure. Further analyses with the whole sample assessed face, criterion-related, and discriminant validity.
\textbf{Results:} EFA supported a two-factor structure (technical and sociotechnical infrastructure), accounting for 65\% of the total variance with strong loadings. CFA confirmed excellent model fit. Face and content validity were supported by the item content reflecting cognitive, affective, and behavioral components. Both subscales were correlated with job satisfaction, perceived autonomy, and feedback from the job itself, supporting convergent validity. Regression analysis supported criterion-related validity, while the Heterotrait–Monotrait ratio of correlations (HTMT), the Fornell–Larcker criterion, and model comparison all supported discriminant validity.
\textbf{Discussion:} The resulting scale is a valid instrument for measuring attitudes toward technical and sociotechnical infrastructure in software engineering research. Our work contributes to ongoing efforts to integrate psychological measurement rigor into empirical and behavioral software engineering research. The scale can also be used by companies and organizations to assess their employees' attitudes toward the infrastructure they provide.
\end{abstract}





\maketitle

\section{Introduction}

%Software engineering research and practice have long been concerned with comparing tools, techniques, and programming languages. For example, the Stack Overflow Developer Survey highlights developers' most admired, desired, and avoided technologies.\footnote{\url{https://survey.stackoverflow.co/2024/technology\#most-popular-technologies}}  However, systematically measuring attitudes toward these technologies remains a challenge. Technologies are often embedded within broader tech stacks, which hindersisolating their individual impact. Moreover, developers' perceptions, such as interest, admiration, or perceived usefulness, are unobservable phenomena that require careful measurement of latent constructs.

Software engineering (SE) researchers are increasingly aware not only of the importance of psychological phenomena (e.g., beliefs, attitudes, intentions) in determining project outcomes, which is central to \textit{behavioral software engineering}~\cite{lenberg2015behavioral}, but also of the difficulty of measuring these phenomena~\cite{ralph2024teaching}. While the importance of scale creation has long been understood in psychology~\cite{jebb2021review}, scale creation has historically been very limited in SE. More recently, however, researchers working in software engineering have awoken to this fact. Graziotin et al.~\cite{graziotin2021psychometrics} call for greater attention to creating and validating psychometric instruments for SE research, including more rigorous and theoretically grounded tools for measuring psychological and behavioral constructs.

In modern software development, professionals rely heavily on the infrastructure supporting their work. This includes software tools, but also processes and norms that shape a work environment. Infrastructure for software development captures technical infrastructure (e.g., CI/CD pipelines, IDEs) as well as sociotechnical infrastructure (e.g., communication tools, documentation standards). However, we currently lack reliable ways to measure the perceptions of these infrastructures, especially in survey-based research. One solution is to assess software professionals' attitudes toward their infrastructure, as attitudes are thought to be good predictors of behavior~\cite{fishbein1977belief, ajzen1985intentions}. Thus, attitudes can act as proxies for infrastructure quality in studies looking at professional beliefs (e.g., job satisfaction) and behavior (e.g., hiring and quitting). We therefore created and validated a psychometric scale to measure developers' attitudes toward technical and sociotechnical infrastructure, guided by the following research question:

\ResearchQuestion{Can software professionals attitudes toward technical and sociotechnical infrastructure be validly measured using a psychometric scale?}

Such an instrument is relevant for both software engineering research and practice. Researchers can use the instrument to consider infrastructure as a factor in their studies with developers. Practitioners can use the instrument in longitudinal developer surveys that many companies and organizations regularly conduct~\cite{DBLP:journals/software/DAngeloLDEHGJ24}.

%In the following, Section~\ref{sec:background} introduces the theoretical basis of attitudes and relevant psychological models, together with a brief overview of software-development-related scales. Section~\ref{sec:method} describes our scale development process, including construct definition, item generation, and refinement through panel review and expert scoring. In Section~\ref{sec:results}, we report on the administration of the scale to a sample of software professionals (N=225). An exploratory factor analysis (EFA) on one half of the randomly split sample revealed strong factor loadings and excellent model fit. We then validated the structure using confirmatory factor analysis (CFA) on the other half of the sample, which, again showed excellent fit. Next, we assessed the scale’s content, face, criterion related, and discriminant validity. Section~\ref{sec:discussion} discusses the implications of our findings and outlines key threats to validity. We conclude the paper in Section~\ref{sec:conclusions}.

%(e.g., RMSR = 0.04, TLI = 0.908)
%(CFI = 0.996, TLI = 0.994)

\section{Background}\label{sec:background}
This section reviews: (1) theories and definitions related to infrastructure; (2) theories and definitions of attitudes; (3) seminal theories from social psychology that explain phenomena using attitudes and their application to software engineering research; (4) measurement and validity-related work in software engineering.

\subsection{Software development infrastructure}\label{ssec:infra}
%What is infrastructure? Why does it matter? What's the difference between technical and sociotechnical? Give examples. What is known about technical and sociotechnical infrastructure for software development? Why should we take an "infrastructure" lens instead of a "tooling" or "practices" lens?

%technical infrastructure: "Technical infrastructure refers to the computer hardware, software, programming languages, frameworks, version control systems, sketching tools, testing tools, networks, security mechanisms, and other technical systems used in day-to-day work.".

Modern software development work is characterized by multitudinous tools, technologies, systems, practices, methods, policies, standards, rituals, and ways of working. In their book \textit{Software engineering at Google}, Winters, Manshreck, and Wright~\cite{winters2020software} devote whole chapters to examples including version control, build systems, and continuous integration. Such collections of systems have been called ``work oriented infrastructures'', due to sharing the same general characteristics as traditional infrastructure~\cite{hanseth2001designing}. 

In information systems and human-computer interaction research, studies focusing on infrastructure have long been common. Particularly, the seminal works by Star~\cite{star1996steps, star1999ethnography} introduce the concept of sociotechnical infrastructure and demonstrate the importance of studying it. While originally introduced to HCI and IS research, these ideas are increasingly applicable to modern software development environments, which depend on layered, intertwined technical and social systems. The popularity of the concept is demonstrated by the reference counts of these articles, as well as the recent systematic review by Lyu et al.~\cite{lyu2025systematic}, which goes over 190 primary studies noting that the main themes investigated in HCI research are growing infrastructure, appropriating infrastructure, and coping with infrastructure. 

Why, then, is studying infrastructure important? The economic theory of infrastructure-led development~\cite{agenor2010theory} posits that public infrastructure drives long term economic growth due to network effects and complementarities, but that the benefits from the investment are non-linear. Such investment in software infrastructure is also done with public money\footnote{https://www.sovereign.tech/programs/fund}. Agenor~\cite{agenor2010theory} uses the example of roads and electricity being needed on one hand to produce commodities in rural areas, but also to transport them to urban areas. Examples for the current time could include how investments into mobile and financial infrastructure enable increased usefulness of them in allowing for network effects from externalities such as application stores, mobile payments, or e-commerce platforms. These complementarities amplify the value of the original infrastructure by enabling new services and innovations that rely on it, increasing and reinforcing adoption and participation. In our view, analogous network effects can be seen in infrastructure used in software development. For example, the investment in version control systems, continuous integration pipelines, or shared libraries allow for the use of them across projects, teams and companies. This in turn allows more developers to use them, rewarding standardizing and documentation efforts and reducing barriers to entry. Complementarities arise with additional tools such as deployment automation, collaborative platforms or the recent trend of integrating LLMs into software development. All of these additions have enhanced the utility of the core infrastructure.

However, it has been argued that infrastructure should be divided into groups when investigated, e.g. an archaeological typology of infrastructure divides infrastructure into static (e.g. storehouse), circulatory (e.g. highway), bounding (e.g. palisade) and signaling (e.g. lighthouses) infrastructure~\cite{wilkinson2019towards}. One way of doing this is by dividing infrastructure according to the function it performs~\cite{buhr2003infrastructure}. However, as we are interested in infrastructure that is used for software development, we argue that these kinds of typologies do not work as same systems can have multiple purposes simultaneously. As an example, a version control system simultaneously allows for asynchronous work on a complex system with internal dependencies, documenting changes, branching and merging features, conflict resolution, and release planning, to name a few. This division would be thus blurry and ambiguous.

To address these challenges, we draw from systems modeling literature, which offers a productive way to conceptualize infrastructure based on its dependence on human actors and institutions. Ottens et al.~\cite{ottens2006modelling} define technical systems as ``systems that perform their function without either actors or social institutions'' or ``systems in which some actors perform subfunctions but social institutions play no role'' and consequently sociotechnical systems as ``engineering systems that need actors and some social/institutional infrastructure to be in place in order to perform their function''. Thus, in an analogous way, we divide technical and sociotechnical infrastructure along these lines. To us, software development systems that professionals use alone are part of technical infrastructure, whereas systems that require multiple actors to perform their function are sociotechnical infrastructure.

\subsection{What are attitudes?}
Attitudes are ``overall evaluations of objects'', which consist of ``affective information (e.g., feelings towards an object), cognitive information (e.g., beliefs associated with an object) and behavioral information (e.g., past experiences with an object)''~\cite{haddock2004contemporary}. This tripartite model of attitudes has held since the '60s~\cite{OSTROM196912}. However, attitudes are more stable over time than core affect~\cite{haddock2004contemporary}, understood as ``neurophysiological state that is consciously accessible as a simple, nonreflective feeling that is an integral blend of hedonic (pleasure–displeasure) and arousal (activation–deactivation) values''~\cite{russell2003core}. This relative stability of attitudes makes them attractive and potentially useful for investigating enduring psychological states toward objects or systems, especially given that much existing research in software engineering has already focused more on transient affective states than on more stable attitudes~\cite{graziotin2015feelings, novielli2019sentiment}.

\subsection{Why measure attitudes?}\label{ss:whyattitudes}
Attitudes are part of several seminal theories in psychology, which are used to explain a multitude of outcomes. These outcomes include, but are not limited to, volitional behavior (actions under conscious control)~\cite{fishbein1977belief}, behavior with limited control~\cite{ajzen1985intentions}, behavior in regard to perceived chances of success and values~\cite{atkinson1957motivational}, acceptance of technology~\cite{davis1989perceived, venkatesh2003user}, cognitive consistency~\cite{heider2013psychology}, attitude change~\cite{kitchen2014elaboration}, identity and self~\cite{blumer1986symbolic, katz1960functional}, emotional regulation~\cite{festinger1957theory, harmon2019introduction}, and perceived values and motivation~\cite{stern1999value, bandura1986social}. 

Three seminal theories referenced above are used to explain human behavior, including the Theory of Reasoned Action (TRA)~\cite{fishbein1977belief}, the Theory of Planned Behavior (TPB)~\cite{ajzen1985intentions}, and Expectancy-Value Theory~\cite{atkinson1957motivational, eccles1983expectancies}. Both TRA and TPB assume that behavior follows from intentions, and intentions from attitudes and social norms. Expectancy-Value Theory~\cite{atkinson1957motivational, eccles1983expectancies} posits that behavior is influenced by beliefs about whether engaging with a task or system (e.g., using a tool or following a process) will lead to desired outcomes (expectancy), and by the value placed on those outcomes. From the point of view of these theories, attitudes toward technical and sociotechnical infrastructure are thus shaped by developers’ beliefs about their usefulness and the importance they attach to outcomes like productivity and collaboration.

Theories such as Cognitive Dissonance Theory~\cite{festinger1957theory} and the Elaboration Likelihood Model~\cite{petty1986elaboration} explain attitude change through inconsistencies or persuasive communication. For example, developers may dislike documentation or testing practices but adjust their views over time. Our scale enables tracking such changes.

Theories like Symbolic Interactionism~\cite{blumer1986symbolic} and Social Cognitive Theory (SCT)~\cite{bandura1986social} highlight how attitudes are shaped through social roles, observation, and perceived efficacy. For example, a developer placed in a mentoring role may adopt more positive views toward documentation tools. Similarly, Value-Belief-Norm Theory (VBN)~\cite{stern1999value} explains how attitudes can underpin moral obligation, motivating developers to act in alignment with shared values.

Thus, measuring developers’ attitudes toward technical and sociotechnical infrastructure would be valuable in survey research for a multitude of reasons. The scale could offer insight into how well work environments support, for example, productivity, collaboration and job satisfaction. This is done by introducing these constructs into a set of related constructs that are often called a nomological networks, i.e. ``interlocking system of laws which constitute a theory''~\cite{cronbach1955construct}.

\subsection{Scales in software engineering research}

A recent book chapter has called attention to measurement issues in software engineering research~\cite{ralph2024teaching}. This is further emphasized with calls for software development specific scale development~\cite{graziotin2021psychometrics}. However, rather than using scales that are developed specifically for software professionals in survey research, it is common to adapt items from scales that have been validated with more general samples (e.g.,~\cite{graziotin2014happy, dutra2022tact}). While some scale development papers exist~\cite{mattos2024instrument}, they have stopped after expert evaluation and do not generally administer the instrument to practitioners, nor do they rigorously and quantitatively assess the validity of the developed measures. To our knowledge, the present work's highly novel contribution to the field of software engineering is in that it documents the development and evaluation of a scale with psychometric properties.
\section{Method}\label{sec:method}

\begin{figure}[t!]
\centering
\includegraphics[width=\linewidth]{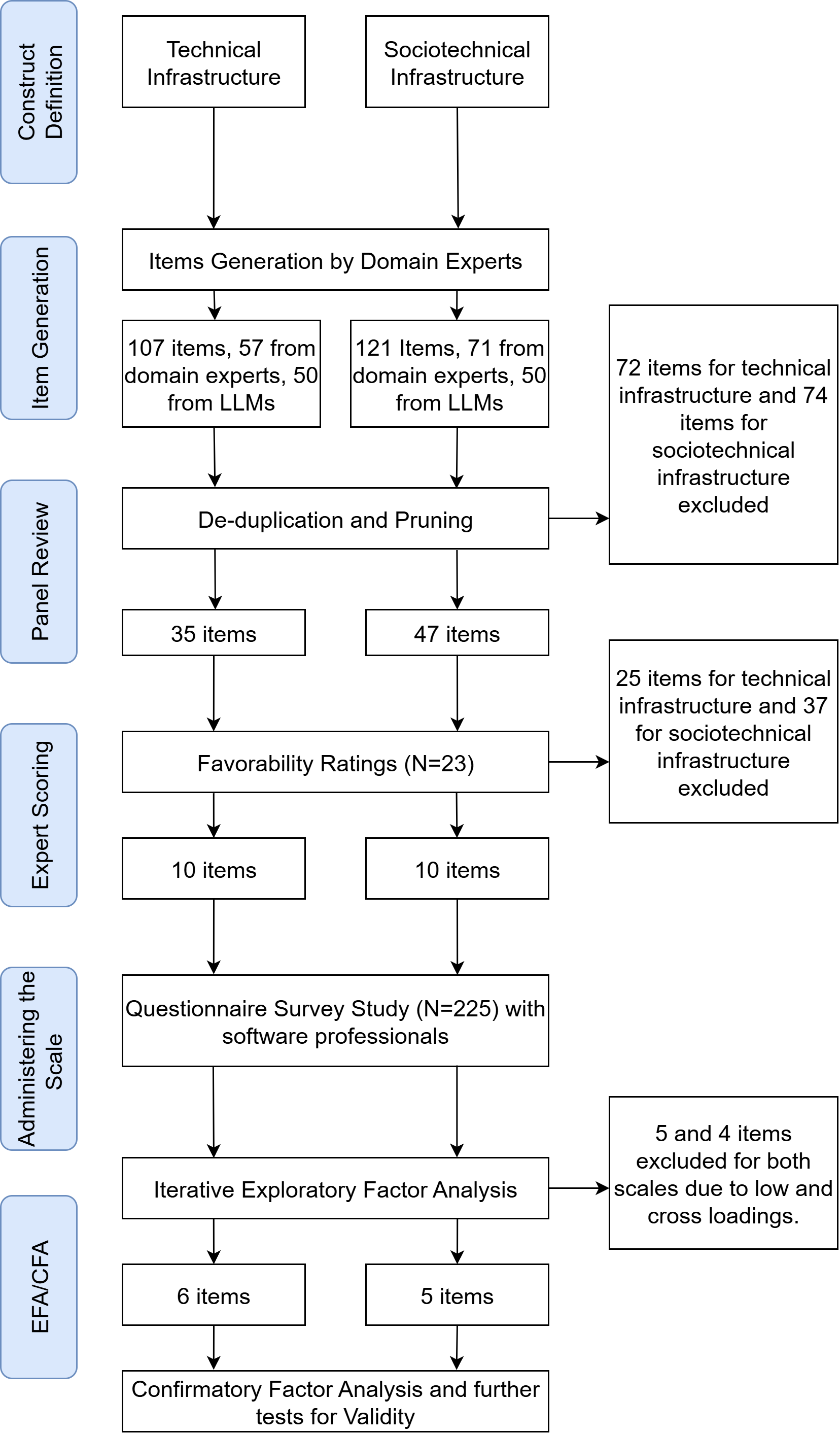}
\caption{PRISMA type diagram showing the steps in creating the scale.}
\Description[The figure shows the development of two scales measuring attitudes toward technical and sociotechnical infrastructure in software development. Items were generated by domain experts and LLMs, then reviewed, rated, and refined through expert scoring and survey data (N=225). Exploratory and confirmatory factor analyses led to final scales with 6 and 5 items respectively.]{Scale development process for measuring attitudes toward technical and sociotechnical infrastructure. The figure outlines the multi-stage procedure used to develop and validate two psychometric scales. First, construct definitions were specified for technical and sociotechnical infrastructure. Items were then generated by domain experts and large language models (LLMs), resulting in 107 and 121 initial items respectively. Through de-duplication and pruning, this was reduced to 35 and 47 items. A panel of 23 raters assessed item favorability, from which the 10 most favorable items for each construct were retained. These items were administered in a survey to 225 software professionals. Iterative exploratory factor analysis (EFA) led to the exclusion of items with low or cross loadings, resulting in a final set of 6 items for technical infrastructure and 5 for sociotechnical infrastructure. These items were then evaluated through confirmatory factor analysis (CFA) and tests for face, convergent and discriminant validity.}
\label{fig:model}
\end{figure}

We followed Trochim's~\cite{trochim2001research} guidelines for developing Likert scales, with reference to relevant sections of Graziotin et al.'s~\cite{graziotin2021psychometrics} guidelines. Trochim~\cite{trochim2001research} recommends a four-step approach: (1) Define the construct being measured, (2) create a pool of possible items to measure the construct, (3) panel review and scoring of items by experts and (4) select a subset of items based on these evaluations. In addition, we (5) administered the scale for a larger audience and assessed reliability and validity. Figure~\ref{fig:model} visualizes this process and shows the number of possible questionnaire items after each step. 

We provide a comprehensive replication package including all of the items and their scoring at each stage (see \nameref{sec:data_availability}). In the following subsections, we elaborate on each step.

\subsection{Defining the constructs}
Both Trochim's~\cite{trochim2001research} and Graziotin's~\cite{graziotin2021psychometrics} guidelines start by defining what we want to measure. To measure software engineers' attitudes toward systems that they use at work, we developed a construct that integrates two complementary dimensions: technical infrastructure and sociotechnical infrastructure.

These dimensions are built on the perspectives offered in Section~\ref{ssec:infra}. Technical infrastructure includes the tangible and primarily automated systems that support software professionals day-to-day development work, such as computer hardware and software, programming languages, frameworks, IDEs, version control systems, sketching and testing tools. These are all tools and systems that are often, but not always, used by a single actor to perform their work duties in software development work. In contrast, sociotechnical infrastructure encompasses systems that need multiple actors and institutional infrastructure to perform their function and allow individual software professionals to coordinate, collaborate and manage software development work. In sociotechnical infrastructure, we include tools, processes, roles, norms, management, and collaborative practices that are used to communicate and organize software development work. The tools included, but not necessarily limited to, are communication and information sharing tools, documentation, coding standards, pair programming and code ownership policies, peer code review practices, sprints, user story planning, and team meetings. This distinction divides software development work into two categories: (1) technical work, where the individual primarily interacts with technical systems to produce technical artifacts, and (2) sociotechnical work, where the individual engages with collaborative systems and practices to coordinate and manage work with others. Thus, we define:

%Prior research in software engineering has traditionally emphasized the technical infrastructure that supports software development~\cite{planning2002economic, hutchinson2021towards}. Accordingly, we define:
%\definition{Technical Infrastructure}{ The tools and technical systems individuals use to perform their work, such as computer hardware, software, programming languages, frameworks, version control systems, sketching tools, testing tools, networks, security mechanisms and other technical systems.}

\begin{definition}[Technical Infrastructure]
%The computer hardware, software, programming languages, frameworks, version control systems, sketching tools, testing tools, networks, security mechanisms, and other technical systems used in day-to-day work.
The tools and technical systems individuals use to perform their work, such as computer hardware, software, programming languages, frameworks, version control systems, sketching tools, testing tools, networks, security mechanisms, and other technical systems.
\end{definition}

%Furthermore, recent research emphasizes the sociotechnical nature of software engineering, emphasizing that social and technical aspects are deeply intertwined and isolating one from the other can lead to an incomplete understanding of engineering practice~\cite{storey2020software}. Building on this view, we define:

\begin{definition}[Sociotechnical Infrastructure] The tools, practices, and processes that are used to communicate and organize work, such as management processes, communication tools, information sharing tools, documentation, coding standards, pair programming, code ownership policies, peer code review practices, sprints, user stories, and team meetings.
\end{definition}

Viewing software development infrastructure through these two dimensions allows researchers to investigate how technical and sociotechnical elements interrelate with each other and allows for analyzing how these dimensions affect factors such as productivity, well-being and team dynamics. These two dimensions also allow for a holistic view beyond individual tools or practices. In our view, this holistic approach allows for measurement that has less construct overlap and contextually confounding factors that can be hard to control for~\cite{westfall2016statistically}. For instance, a single treasured tool could affect measurements of tools that are often used alongside it in ways reminiscent of the halo effect~\cite{nisbett1977halo}. Thus, by dividing the concepts this way with the still broad and holistic concepts of technical and sociotechnical infrastructure, software engineering research can capture software professionals' attitudes and experiences.

%Together, these two definitions form the basis of what we want to measure.

%What I will write about here: distinction between between technical and sociotechnical grounded in previous research (see new 2.1)- Ontological argument?

%Write about: automation vs coordination, artifacts vs practices. These maps into our concepts.

%We have more than one because we want to investigate how they relate to each other: how practices shape tool use etc. We dont have more than two because we dont want conceptual overlaps that risk confounding effects. Lastly, model parsimony?

\subsection{Generating items}
Trochim calls this step ``Generating the Items''~\cite{trochim2001research}, while Graziotin et al.~\cite{graziotin2021psychometrics} call it ``Pool Items''. We gave eight domain experts preliminary definitions of the two dimensions above and asked them to generate items. By domain expert, we mean individuals with theoretical knowledge and practical experience in the subject, as is common in psychological literature~\cite{haynes1995content}. The eight individuals included a professor, postdoctoral researcher, doctoral students, and a master’s student, all of whom had years of experience in SE research, teaching, and practice. The group of domain experts consisted of individuals with diverse backgrounds, including participants originally from North America, South America, the Caribbean, Africa, the Middle East, and the Nordics. Experts were asked to generate 10 items per dimension, though they could stop at any time. They collectively generated 57 items for technical infrastructure and 71 items for sociotechnical infrastructure. After these preliminary items were collected, the first and second authors revised them for clarity. We have marked the original items and the modifications in our replication package. 

To supplement these pools of items, we generated 50 additional items on the 7th of July 2024 for each type of infrastructure using ChatGPT 4o with the default parameters accessed via a browser, and with various prompting techniques. We used true random numbers from \url{random.org} to choose examples for the prompts that require examples (One-Shot, Few-Shot, and Few-Shot chain of thought) from the list of items generated by domain experts. This process, together with the exact prompts, is documented in our replication package. We did not use LLMs in any other part of this research.

\subsection{Panel review and expert scoring}
Trochim calls this step ``Rating the items'' and recommends it \textit{before} pilot testing~\cite{trochim2001research}, while Graziotin et al. call it ``Item analysis'' and recommend it \textit{after} pilot testing~\cite{graziotin2021psychometrics}. We followed Trochim to limit the workload for the experts providing the  favorability ratings (examples below).

The first two authors pruned and de-duplicated the items in a face-to-face collaborative session. First, both authors read the list of items, and then removed items that were too specific. For example, ``I feel confident in the security measures of our technical infrastructure'' is too specific because it's about security in particular. In contrast, some items were too specific in that they applied only to a subset of software professionals, such as developers working in distributed environments. Next, we grouped similar items and selected the most concise and general item from each group, dropping the others. For example, we chose the item socio2 (see Table~\ref{tab:efa}) over items such as ``I find the current infrastructure supports seamless information sharing'' and ``I find it challenging to share information with the existing tools and processes''. This resulted in a pool of 35 items for technical infrastructure and 47 for sociotechnical infrastructure. We provide a reason for every item dropped at this stage in our replication package.

Next, 23 domain experts scored each item's ``favorability'' toward the target concept. Favorability here does not refer to what the expert believes is the case for them, but rather they are judging ``how favorable each item is with respect to the construct of interest''~\cite{trochim2001research}. The domain experts included 14 university professors, 2 postdocs and 7 graduate students, all working in the software engineering domain. Domain experts received instructions, together with examples, on scoring favorability of each item with regard to the construct it attempts to measure, and the instruction to use a rating scale from +5 to -5. The order of the items being scored was randomized for each rater. We also added three attention check items for both scales to the favorability ratings. These attention check items were either negative statements framed as positive or vice versa, for example: ``Our sociotechnical infrastructure makes workflow so thorough and rigorous that our deliverables are often late''. After a brief trial run, 15 of the experts scored 94.4\% of these items correctly, that is, positive statements as favorable and vice versa. The high rate of correct scoring indicates that the experts paid attention to the items when scoring them. 

\subsection{Selecting items for the preliminary scale}
From the favorability ratings, we calculated total-item correlations as advocated by Trochim~\cite{trochim2001research} and performed a Welch's t-test between the lowest and highest quartiles for assessing discriminative power between items. We observed that two raters scored the technical infrastructure items substantially lower than all others, with total scores of 15 and 29 compared to a range of 77 to 154 for the remaining raters. Due to this large discrepancy, which suggested a misunderstanding of the task or scale, we excluded their ratings from the analysis. As a result, selection scores were calculated using ratings from 21 raters for technical infrastructure and 23 for sociotechnical infrastructure. 

To select the items for the scale, we first removed all of the items with an item-total correlation below Trochim's recommended cut-off of 0.6~\cite{trochim2001research}. Then, for each infrastructure dimension, we selected the 10 items with the highest t-test scores. We then reversed some items so that both dimensions would have 5 regularly scored and 5 reverse-scored items, for example, the item ``Our technical infrastructure is well-maintained'' was changed to ``Our technical infrastructure is poorly maintained''. Our aim here was to have balanced scales for both dimensions, to reduce response biases, to control for acquiescence (i.e. ``yeah saying''), and to disrupt nonsubstantive responding (straightlining, that is, selecting the same option across items)~\cite{weijters2012misresponse}. Following recommendations in the literature~\cite{choi2004catalog}, we avoided negations such as “not” and other modifications making items awkward, lengthy, or difficult to interpret.

\begin{table}[t]
\centering
\small
\caption{Descriptive Statistics for Scale Items. }
\begin{tabular}{l|rrrrr}
\toprule
\textbf{Item ID} & \textbf{Mean} & \textbf{SD} & \textbf{Skewness} & \textbf{Kurtosis} & \textbf{Miss \%} \\
\midrule
tech1  & 3.11 & 1.12 & -0.05 & -0.97 & 2.67\\
tech2  & 2.64 & 1.17 & 0.52 & -0.75 & 0.89\\
tech3  & 2.70 & 1.20 & 0.42 & -0.86 & 2.67\\
tech4  & 3.66 & 0.94 & -0.58 & -0.06 & 1.78\\
tech5  & 3.56 & 1.00 & -0.71 & 0.06 & 0.89\\
tech6  & 2.47 & 1.13 & 0.50 & -0.60 & 1.33\\
tech7  & 3.80 & 0.86 & -0.83 & 0.95 & 0.44\\
tech8  & 3.77 & 1.00 & -0.71 & 0.06 & 0.89\\
tech9  & 3.91 & 0.84 & -1.01 & 1.37 & 0.89\\
tech10  & 2.24 & 1.10 & 0.80 & -0.12 & 0.00\\
socio1  & 2.96 & 1.09 & 0.05 & -0.97 & 2.67\\
socio2  & 3.70 & 0.99 & -0.71 & 0.08 & 0.00\\
socio3  & 2.82 & 1.15 & 0.31 & -0.85 & 0.44\\
socio4  & 2.66 & 1.10 & 0.63 & -0.51 & 1.33\\
socio5  & 3.56 & 1.01 & -0.56 & -0.13 & 2.22\\
socio6  & 3.58 & 0.95 & -0.72 & 0.32 & 0.44\\
socio7  & 3.60 & 1.03 & -0.67 & -0.06 & 0.44\\
socio8  & 2.71 & 1.05 & 0.46 & -0.64 & 0.44\\
socio9  & 3.51 & 0.97 & -0.68 & 0.06 & 4.00\\
socio10 & 2.37 & 0.96 & 0.80 & 0.15 & 0.00\\
\bottomrule
\end{tabular}
\label{tab:descriptives}
\end{table}

\subsection{Assessing the validity of the scale}

While Trochim’s~\cite{trochim2001research} original guidelines suggest that the scale can now be used in practice, contemporary psychometric standards emphasize the importance of empirically evaluating the scale’s factor structure and validity through data collection and analysis~\cite{worthington2006scale, devellis2021scale}.

To administer the newly developed scales, we incorporated them into a broader questionnaire used in a larger survey study~\cite{anon2025}, in which software professionals’ attitudes toward their work infrastructure serves as an exogenous variable. The questionnaire also included related constructs, such as job satisfaction (measured using the \textit{5-item Short Index of Job Satisfaction}~\cite{judge2000personality}) and job quality (captured through the autonomy and feedback from the job itself dimensions of the revised \textit{Job Diagnostic Survey}~\cite{idaszak1987revision}). As shown later in Section~\ref{sec:results}, including these related constructs allows us to rigorously evaluate the proposed scales in terms of both criterion-related and discriminant validity.

The target population for the questionnaire survey where the scale was administered is software professionals, e.g., programmers, testers, technical artists, and their managers, who are primarily employed by a single organization from which they receive the majority of their income, and who work within the software industry. Freelancers and self-employed individuals were excluded from the study. Only respondents who reported being between the ages of 18 and 65 were included, as these were considered the conventional working-age population.

The questionnaire survey was translated from English into Portuguese, Chinese, Thai, Finnish, Swedish, Arabic, Farsi, Urdu, German, Spanish, Japanese, Russian and Turkish, and implemented using \textsf{Opinio}.\footnote{https://www.objectplanet.com/opinio/} Each translation was localized by a researcher with prior experience in conducting human factors research in software engineering. After receiving permission from the Dalhousie research ethics board, the study was advertised in these locales using social media and company contacts (e.g. LinkedIn, internal company communication channels). The responses were received between December 12, 2024 and June 7, 2025. Incentives such as raffles and gift cards were used depending on the locale, however, caution was used so that gift cards were used only when advertising in closed spaces such as company channels. Our recruiting strategy thus corresponds to convenience and breadth sampling~\cite{baltes2022sampling}.

To improve data quality, all demographic questions were optional. We also included an ``I don't know'' option for each scale item~\cite[see][]{oppenheim2000questionnaire,dolnicar2014including} and an attention-check item for the Job Diagnostic Survey~\cite{kung2018attention}. Items scored as ``I don't know'' were coded as missing, while surveys with a failed attention-check item were discarded. The order of the items was randomized to reduce order effects~\cite{malhotra2008completion}. The item-level descriptives of the responses can be seen in Table~\ref{tab:descriptives}. Item means ranged from 2.24 to 3.80, standard deviations from 0.84 to 1.20, all of which are within acceptable bounds. Skewness and kurtosis values also indicated no substantial deviations from normality (all skewness values <|1.1| and all kurtosis values <|1.4|). Item-level missingness was mostly below 2.7\%. We used listwise deletion in EFA and full information maximum likelihood~\cite{enders2001relative} in CFA.

%Things to add here:
%\begin{itemize}
%    \item Study population done
%    \item Selection criteria done
%    \item Sampling approach / recruiting method done
%    \item *brief* description of survey translation, localization, validation and local recruiting, incentives - done
%    \item note survey was implemented using Opinio done
%    \item Note that scale items are always %optional; Planned approach for missing data done
%    \item Survey start date; cut-off data for data reported in this study done
%    \item mention that the study was approved by our REB. If you put the approval letter in supplementary materials, remember to redact all the fields that would identify us or our university
%\end{itemize}

%Sample size
The sample size needed for both EFA and CFA is debated, but the often-cited work by Worthington and Whittaker~\cite{worthington2006scale}, states ``smaller sample sizes may be adequate\dots with at least $4:1$ items per factor and factor loadings greater than $|.6|$'', and recommends a minimum sample size of 100. Our work satisfies these conditions, with an item to response ratio of 5.65 and a sample size of 122 for the EFA. We split our sample randomly using the base R function sample, resulting in 122 and 123 responses used in EFA and CFA, respectively. For later analyses, we used the whole sample.

To explore and improve the underlying structure of the scales and reduce the number of items, we conducted an exploratory factor analysis (EFA) using the maximum likelihood extraction and oblimin rotation with the R package \textsf{psych}~\cite{revelle2015package} (version 2.5.3). Rotation is applied after extraction to improve interpretability by simplifying the factor structure. We chose oblimin rotation (an oblique method) because the underlying constructs we are attempting to measure should be related, and oblimin rotation allows factors to correlate with one another~\cite{fabrigar1999evaluating}. Improving factor structure involves iterating among (1) investigating the factor structure using EFA; (2) dropping items with low communality; (3) dropping items with substantial cross-loadings. The analysis continues until all remaining items load highly on the factor they are supposed to load on, and no others. This process is a powerful tool for assessing and enhancing validity because it simultaneously assesses convergent and discriminant validity. However, it can lead to overfitting, which we mitigate by randomly splitting our dataset in two, using half the data for this exploratory analysis, and the other half for confirmatory analysis.

The exploratory analysis was performed on a subset of 122 valid responses randomly selected from the entire data set. The initial EFA suggested a three-factor solution based on eigenvalues > 1 and parallel analysis, with a cumulative variance explained of 56\%. However, several items showed substantial cross-loadings or weak factor loadings (< .40), and some exhibited high complexity or low communalities (< .30), indicating poor fit. Based on these criteria, we iteratively removed items to improve the clarity of the factor structure. After the final EFA was performed, we assessed the suitability of the data for factor analysis by performing Bartlett’s Test of Sphericity and calculating the Kaiser-Meyer-Olkin (KMO)~\cite{williams2010exploratory} measure using the R package \textsf{psych}~\cite{revelle2015package} (version 2.5.3). This factor structure from the EFA was retained for the rest of the analyses.

To confirm the factor structure, we performed confirmatory factor analysis with the R package \textsf{lavaan}~\cite{rosseel2012lavaan} (version 0.6-18). We performed two confirmatory factor analyses: (1) a model containing only technical and sociotechnical infrastructure factors identified in the EFA and (2) a model including additional factors of job satisfaction~\cite{judge2000personality}, as well as autonomy and feedback from the revised job characteristics model~\cite{idaszak1987revision}. The first model validates the factor structure obtained from the EFA, whereas the second model provides us with factor scores for external constructs to the scale being developed, which were used in later analyses for convergent, criterion-related, and discriminant validity.

We used Job Satisfaction, Autonomy and Feedback from the job itself in later analyses for convergent, criterion-related, and discriminant validity. Job satisfaction is an overall evaluation of a job, and thus we presume it would positively correlate with positive attitudes towards technical and sociotechnical infrastructure. In the job characteristic model introduced by Oldham and Hackman~\cite{hackman1976motivation}, autonomy refers to the degree to which the job provides freedom and discretion to the individual in scheduling their work and in determining the procedures to be used in carrying it out, while feedback refers to the degree to which carrying out work activities results in the employee obtaining direct and clear information about the effectiveness of their performance in the job. Thus, we presume that jobs that are perceived as autonomous and providing feedback in themselves are also on average jobs where technical and sociotechnical infrastructure allow for these through, for example, tooling and job procedures. Hence, we assume that these two constructs are positively associated with higher evaluations of technical and sociotechnical infrastructure.

To assess the criterion-related validity of the newly developed scale, we correlated its factor scores with established constructs of job satisfaction, autonomy and feedback from the job itself using the whole sample. Correlations were calculated using the R package psych~\cite{revelle2015package} version 2.5.3 and regression analysis was performed with the function lm from base R version 4.3.3.

Lastly, to assess the discriminant validity of our scale, we calculated two commonly reported measures, the Heterotrait-Monotrait Ratio of Correlations (HTMT)~\cite{henseler2015new} and the Fornell–Larcker criterion~\cite{fornell1981evaluating}. HTMT was calculated using the function htmt from the R package \textsf{semTools} version 0.5.7. Latent variable correlations used for the Fornell–Larcker discriminant validity analysis were obtained from the CFA model-implied correlation matrix (i.e., function inspect(fit, "cor.lv")). These account for measurement error and reflect the structural relationships between constructs. However, due to recent critique of these measures by R\"onkk\"o \& Cho~\cite{ronkko2022updated}, we also assessed discriminant validity by fitting a baseline CFA model with both dimensions of our infrastructure scale, and fitting an alternative model where the correlation between the constructs was set to 1.0. We compared the model fits with the chi-square test, as well as with the Bayesian information criterion (BIC) and the Akaike Information Criterion (AIC)~\cite{vrieze2012model}.

\section{Results}\label{sec:results}

In this section, we first report the demographics of the respondents to our questionnaire study, next we explore the factor structure of the administered scale, and then we confirm it with a split sample approach. Next, we discuss the face and content validity of the scale by analyzing the items remaining after factor analysis. In the last three subsections, we provide evidence for convergent, criterion-related, and discriminant validity of the developed scale.

%Consider rewriting like this: First we assess convergent and discriminant validity and iteratively refine the scale to improve it. This is exploratory, hence EFA. Next we need to show that the scale still loads right on some other data (mitigate overfitting), so we do CFA. Then, once we're confident that the scale holds (loadings still come out well, good convergent and discriminant validity), then we assess face and content validity. So name the sections after what we are trying to achieve, not the statistical approach. Note that we already assessed face validity while selecting items, but we come back to it with the final version, and we need to reassess content validity because there are fewer items now. In a broad sense, we're arguing that attitude toward infrastructure has two related but unequal / separate subdimensions (technical and sociotechnical) and content validity is good because we have >2 items for each subdimension. But here we are also making sure that we're not just asking five semantically equivalent questions for each topic. We want to make sure each question has its own meaning. Give examples of semantically identical questions like "I like our infrastrucure, our infrastructure is good, our infrastruture works well for me, etc." 
\subsection{Respondent demographics}

\begin{table}[!htb]
\centering
\caption{Self-reported demographic information of the respondents.}
\begin{tabular}{llrr}
\toprule
& \textbf{Demographic Item} & \textbf{N} & \textbf{\%} \\
\midrule
\multirow{3}{*}{\textit{Gender}} 
    & Man & 157 & 70\% \\
    & Woman & 59 & 26\% \\
    & NA / Prefer not to say & 9 & 4\% \\
\midrule
\multirow{5}{*}{\begin{tabular}[c]{@{}l@{}}\textit{Country of}\\\textit{Residence}\end{tabular}} 
    & Brazil & 60 & 27\% \\
    & People's Republic of China & 31 & 14\% \\
    & Finland & 25 & 11\% \\
    & Thailand & 21 & 9\% \\
    & Saudi Arabia & 13 & 6\% \\
\midrule
\multirow{6}{*}{\textit{Survey Language}} 
    & Portuguese & 61 & 27\% \\
    & Chinese & 48 & 21\% \\
    & Thai & 27 & 12\% \\
    & Finnish & 23 & 10\% \\
    & English & 21 & 9\% \\
    & Other & 45 & 20\% \\
\midrule
\multirow{6}{*}{\begin{tabular}[c]{@{}l@{}}\textit{Age}\\\textit{(Range: 19--61)}\end{tabular}} 
    & Mean: 35.4, Median: 34 & & \\
    & 18--24 & 13 & 6\% \\
    & 25--30 & 66 & 29\% \\
    & 31--40 & 87 & 39\% \\
    & 41--50 & 39 & 17\% \\
    & 55 or more & 20 & 9\% \\
\midrule
\multirow{5}{*}{\begin{tabular}[c]{@{}l@{}}\textit{Years of}\\\textit{Experience}\\\textit{(Range: 0--40)}\end{tabular}} 
    & Mean: 10.4, Median: 8 & & \\
    & 0--4 years & 57 & 25\% \\
    & 5--9 years & 63 & 28\% \\
    & 10--19 years & 58 & 26\% \\
    & 20+ years & 33 & 15\% \\
\midrule
\multirow{4}{*}{\textit{Highest Education}} 
    & Doctorate & 15 & 7\% \\
    & Master's & 80 & 36\% \\
    & Bachelor's & 93 & 41\% \\
    & Other or NA & 37 & 16\% \\
\midrule
\multirow{7}{*}{\textit{Company Size}} 
    & 0--9 & 11 & 5\% \\
    & 10--99 & 46 & 20\% \\
    & 100--999 & 42 & 19\% \\
    & 1000--9999 & 69 & 31\% \\
    & 10,000--99,999 & 23 & 10\% \\
    & 100,000 or more & 14 & 6\% \\
    & NA & 20 & 9\% \\
\bottomrule
\end{tabular}
\begin{tablenotes}
\small
\item All demographic questions but age were optional to answer.
\end{tablenotes}
\label{tab:demo}
\end{table}

We received complete responses from 239 software professionals. After removing answers with failed attention check items, we were left with a total of 225 answers. We report demographics in Table \ref{tab:demo}. Note that none of the demographic questions other than age were mandatory to answer, meaning the number of responses from question to question changes. Respondents reported a wide range of software development experience (0–40 years, mean = 10.4), with the majority holding a bachelor's (41\%) or master’s degree (36\%). The participants were employed in organizations of varying sizes, with nearly one-third working in companies with 1,000–9,999 employees. Respondents were from diverse geographic and linguistic backgrounds, with Brazil, China, and Finland being the most common countries of residence, and Portuguese, Chinese, and Thai among the most common languages the survey was taken in.

\subsection{Exploratory factor analysis}
Following item reduction, at the end of the iterative EFA process, we fitted an EFA model using maximum likelihood extraction with oblimin rotation on a refined set of 11 items (see Table~\ref{tab:efa}). The results supported a two-factor structure consistent with our theoretical distinction between technical (6 items) and sociotechnical infrastructure (5 items). Items showed strong primary loadings (ranging from .66 to .88), minimal cross-loadings, and high communalities ($h^2 \ge .55$). The two factors were moderately correlated ($r = .63$), supporting the use of an oblimin rotation.

The chi-square test of model fit was statistically significant, $\chi^2(34) = 56.75$, $p = .009$, which is not uncommon given the sample size ($N = 122$). However, alternative fit indices indicated excellent model fit (RMSEA = .077, 90\% CI [.039, .112]; RMSR = .03; TLI = .953), supporting the adequacy of the two-factor solution. The two factors accounted for 65\% of the total variance, and item complexity was low across the board (mean = 1.0). This 11-item structure was retained for confirmatory factor analysis using a split sample.

We performed post hoc tests for sampling adequacy for the data used in the EFA, using the Kaiser-Meyer-Olkin (KMO) measure~\cite{williams2010exploratory}. The overall measure of sampling adequacy (MSA) provided by the test was 0.90, which corresponds to marvelous adequacy for factor analysis in the original Kaiser's evaluation levels~\cite{kaiser1974index}. All individual item MSAs were above 0.85, ranging from 0.86 to 0.93, supporting the inclusion of all retained items. We also performed Bartlett’s Test of Sphericity, which was significant \((\chi^2(45) = 825.59, p < .001)\), indicating that the correlations among items were sufficiently large for factor analysis.

\begin{table*}[ht]
\small
\caption{Items after the iterative exploratory factor analysis process.}
\centering
\small
\begin{tabular}{l|p{8.5cm}|ccccc|c}
\hline
ID & Item & F1 & F2 & Comm. & Uniq. & Comp. & MSA \\
\toprule
tech4 & The technical infrastructure I use at work is well made & 0.88 & 0.00 & 0.78 & 0.22 & 1.0 & 0.91 \\
tech5 & I am satisfied with the technical infrastructure provided at my workplace & 0.88 & -0.02 & 0.75 & 0.25 & 1.0 & 0.91 \\
tech7 & Our technical infrastructure works well & 0.82 & 0.09 & 0.77 & 0.23 & 1.0 & 0.93 \\
tech8 & The technical infrastructure I use helps me work more efficiently & 0.71 & 0.13 & 0.64 & 0.36 & 1.0 & 0.87 \\
tech9 & Our infrastructure includes tools appropriate for my work tasks & 0.73 & 0.13 & 0.64 & 0.36 & 1.1 & 0.93 \\
tech10 & The technical infrastructure at my workplace is unreliable & 0.79 & -0.07 & 0.56 & 0.44 & 1.0 & 0.91 \\
\midrule
socio2 & Sociotechnical infrastructure makes it easy to share information & -0.05 & 0.83 & 0.64 & 0.36 & 1.0 & 0.89 \\
socio5 & Sociotechnical infrastructure at work is user-friendly & 0.15 & 0.66 & 0.58 & 0.42 & 1.1 & 0.89 \\
socio6 & Our workflow is enhanced by the sociotechnical systems & 0.06 & 0.77 & 0.65 & 0.35 & 1.0 & 0.89 \\
socio7 & Sociotechnical infrastructure encourages learning and improvement & -0.09 & 0.79 & 0.55 & 0.45 & 1.0 & 0.86 \\
socio9 & Sociotechnical infrastructure supports different team needs & 0.04 & 0.85 & 0.78 & 0.22 & 1.0 & 0.86 \\
\bottomrule
\end{tabular}
\begin{tablenotes}
\small
\item \textbf{Model fit:} RMSEA = 0.077 (90\% CI: 0.039–0.112), RMSR = 0.03, TLI = 0.953
\end{tablenotes}
\label{tab:efa}
\end{table*}

\subsection{Confirmatory factor analysis}
Construct validity refers to ``the degree to which a test or instrument is capable of measuring a concept, trait, or other theoretical entity.''~\cite{apa_dictionary}. We conducted a confirmatory factor analysis (CFA) to evaluate the factorial validity of a two-factor model with a separate sample. This was done with the factor structure identified by the EFA described in the previous subsection. Model fit indices are shown in Table~\ref{tab:cfa_fit} and the loadings in Table~\ref{tab:cfa_loadings}. The model included 11 items loading onto two correlated latent constructs: technical infrastructure and sociotechnical infrastructure. All items loaded significantly (p < .001) on their respective factors, with standardized loadings ranging from 0.588 to 0.853 for technical infrastructure and 0.685 to 0.866 on sociotechnical infrastructure. One item in the technical infrastructure construct exhibited a slightly lower loading (0.588), which we decided to retain due to negatively worded items being helpful for lowering response bias.

\begin{table}[ht]
\small
\centering
\caption{Standardized Factor Loadings from CFA Model}
\begin{tabular}{llc}
\toprule
\textbf{Factor} & \textbf{Item} & \textbf{Std. Loading} \\
\midrule
\multirow{5}{*}{\textbf{Technical Infrastructure}} 
& tech4 & 0.831 \\
& tech5 & 0.845 \\
& tech7 & 0.853 \\
& tech8 & 0.764 \\
& tech9 & 0.754 \\
& tech10 & 0.588 \\
\midrule
\multirow{5}{*}{\textbf{Sociotechnical Infrastructure}} 
& sociotech2 & 0.701 \\
& sociotech5 & 0.804 \\
& sociotech6 & 0.866 \\
& sociotech7 & 0.685 \\
& sociotech9 & 0.853 \\
\bottomrule
\end{tabular}
\label{tab:cfa_loadings}
\end{table}

%\begin{table*}[ht]
%\centering
%\small
%\caption{Fit indices for confirmatory factor analysis models}
%\begin{tabular}{lcc}
%\toprule
%& \textbf{Two-Factor Model (Tech + Sociotech)} & \textbf{Five-Factor Model (All Constructs)} \\
%\textbf{Sample Used} & (CFA Split) & (Whole Sample) \\
%\midrule
%\textbf{Fit Index} & & \\
%$\chi^2$ (df), $p$         & 48.1 (43), $p = .275$     & 297.05 (199), $p < .001$ \\
%CFI                        & 0.993                      & 0.962 \\
%TLI                        & 0.991                      & 0.956 \\
%RMSEA [90\% CI]            & 0.032 [0.000, 0.074]       & 0.047 [0.035, 0.058] \\
%SRMR                       & 0.041                      & 0.052 \\
%\bottomrule
%\end{tabular}
%\begin{tablenotes}
%\small
%\item \textit{Note.} The two-factor model includes only the technical and sociotechnical infrastructure constructs. %The five-factor model includes those two plus autonomy, feedback, and job satisfaction.
%\end{tablenotes}
%\label{tab:cfa_fit}
%\end{table*}

\begin{table}[ht]
\centering
\small
\caption{Fit indices for CFA models}
\label{tab:cfa_fit}
\begin{tabular}{@{}lcc@{}}
\toprule
\textbf{Fit Index} & \textbf{Two-Factor} & \textbf{Five-Factor} \\
 & (CFA Split) & (Full Sample) \\
\midrule
$\chi^2$ (df), $p$ & 48.1 (43), $p = .275$ & 297.05 (199), $p < .001$ \\
CFI                & 0.993                 & 0.962 \\
TLI                & 0.991                 & 0.956 \\
RMSEA [90\% CI]    & 0.032 [0.000, 0.074]  & 0.047 [0.035, 0.058] \\
SRMR               & 0.041                 & 0.052 \\
\bottomrule
\end{tabular}
\vspace{1mm}
\begin{minipage}{\linewidth}
\footnotesize \textit{Note.} Two-factor = technical + sociotechnical infrastructure. Five-factor adds autonomy, feedback, and job satisfaction.
\end{minipage}
\end{table}

The model demonstrated excellent fit to the data (N = 123): $\chi^2$(43) = 48.062, p = .275, CFI = 0.993, TLI = 0.991, RMSEA = 0.032, 90\% CI [0.000, 0.074], with p = .715 for RMSEA $\le$ .05, SRMR = 0.041. The two latent constructs were moderately correlated (standardized estimate = 0.601, p < .001), supporting their conceptual relatedness while retaining discriminant validity.

\subsection{Face and content validity}\label{ssec:facevalidity}

Content validity refers to ``the extent to which a specific set of items reflects a content domain''~\cite{devellis2021scale}. As discussed in the background section, attitudes are overall evaluations of objects that comprise three types of information: cognitive, affective, and behavioral~\cite{haddock2004contemporary}. These correspond to beliefs, feelings, and past experiences, respectively. Examining the results of the EFA (Table~\ref{tab:efa}), we find that the items selected through our scale development for both the technical and sociotechnical infrastructure reflect all three components.

For the technical infrastructure items, tech4 and tech7 primarily convey cognitive information, expressing beliefs about the infrastructure being well made and functioning effectively. Item tech5 reflects affective content by referring to satisfaction with the infrastructure. Items tech8, tech9, and tech10 incorporate behavioral information, as they refer to personal experiences using the infrastructure to perform tasks efficiently and with appropriate tools.

The sociotechnical items similarly represent the tripartite structure. Items socio2 and socio5 contain cognitive content, describing beliefs about ease of information sharing and user-friendliness. Items socio6 and socio9 reflect behavioral experiences related to enhanced workflows and responsiveness to diverse team needs. Finally, item socio7 includes an affective element, describing the infrastructure as encouraging continuous learning and improvement.

%In summary, the selected items for both dimensions capture a broad range of content and align well with the theoretical structure of attitudes, showing support for the face validity of the constructs.

\subsection{Convergent validity}
% Correlations with job satisfaction and autonomy, feedback of job characteristics
``Convergent validity is evidence of similarity between measures of theoretically related constructs''~\cite{devellis2021scale}. To examine convergent validity of the new scale, we computed Pearson correlations between the factor scores of two dimensions, Tech and Sociotech, and factor scores of relevant theoretically related constructs: perceived autonomy, feedback from the job itself, and job satisfaction (see Table~\ref{tab:corr}). As expected, Tech and Sociotech dimensions were strongly correlated with each other (r = .68, p < .001), supporting the idea that they capture related but distinct aspects of attitudes toward work infrastructure.

Both dimensions also showed positive correlations with job-related attitudes. Specifically, technical infrastructure was moderately associated with job satisfaction (r = .47, p < .001), feedback from the work itself (r = .41, p < .001), and autonomy (r = .34, p < .001). Sociotechnical infrastructure was likewise significantly correlated with job satisfaction (r = .39, p < .001), feedback (r = .48, p < .001), and autonomy (r = .27, p < .01). These results indicate that more favorable attitudes toward technical and sociotechnical infrastructure are associated with greater perceived autonomy, better perceived feedback from the work itself, and higher job satisfaction, supporting the convergent validity of the scale.

\subsection{Criterion-related validity}
%Regression on job satisfaction

For a scale to have criterion (or criterion-related) validity, it needs to be have empirical associations with some criterion, whether or not that association is understood theoretically~\cite{devellis2021scale}. In this study, we assessed concurrent validity, a subtype of criterion-related validity, by testing whether attitudes toward technical and sociotechnical infrastructure were associated with job characteristics and satisfaction measured at the same time. Specifically, we used a series of linear regression models to examine whether these attitudes predicted autonomy, feedback from the job itself, and job satisfaction. The results can be seen in Table~\ref{tab:regression_results}. Technical infrastructure significantly predicted autonomy ($\beta = .31$, $p = .001$), while sociotechnical infrastructure did not. Both types of infrastructure predicted feedback from the job itself, with sociotechnical infrastructure showing a notably stronger effect ($\beta = .49$, $p < .001$). In predicting job satisfaction, technical infrastructure was a robust predictor both directly ($\beta = .57$, $p < .001$) and after adding autonomy and feedback to the model ($\beta = .41$, $p < .001$). Feedback was the strongest predictor of job satisfaction in the full model ($\beta = .60$, $p < .001$), while the effect of sociotechnical infrastructure dropped to non-significance, suggesting that its influence is mediated through feedback mechanisms.

%\end{table*}

\begin{table}[t]
\centering
%\small
\caption{Correlations between latent factor scores.}
\label{tab:corr}
\begin{tabular}{lccccc}
\toprule
 & Tech. & Socio. & Auto. & Feed. & Sat. \\
\midrule
Tech.     & 1.00 & .68$^{***}$ & .34$^{***}$ & .41$^{***}$ & .47$^{***}$ \\
Sociotech    &      & 1.00        & .27$^{***}$ & .48$^{***}$ & .39$^{***}$ \\
Autonomy     &      &             & 1.00        & .83$^{***}$ & .57$^{***}$ \\
Feedback     &      &             &             & 1.00        & .64$^{***}$ \\
Satisfaction      &      &             &             &             & 1.00        \\
\bottomrule
\end{tabular}
\vspace{3pt}
\begin{minipage}{\linewidth}
\scriptsize \textit{Note.} Tech = Attitude towards technical infrastructure, Socio = Attitude towards sociotechnical infrastructure, Auto = Autonomy, Feed = Feedback, Sat = Job Satisfaction. $^{***}p < .0001$.
\end{minipage}
\end{table}

\begin{table}[t]
\small 
\centering
\caption{Four linear regression models predicting autonomy, feedback, and job satisfaction with attitudes toward technical and sociotechnical infrastructure. Estimate followed by the $p$-value in parenthesis in the line below}
\begin{tabular}{lcccc}
\toprule
\textbf{Predictor} & \textbf{Autonomy} & \textbf{Feedback} & \multicolumn{2}{c}{\textbf{Job Satisfaction}}  \\
& & & \textbf{Direct} & \textbf{Mediated} \\
\midrule
\textbf{Tech}        & 0.310                & 0.192               & 0.573                &  0.416 \\
$p$                  & (\textbf{$<$0.0001}) & (\textbf{0.045})    & (\textbf{$<$0.0001}) & (\textbf{0.0001})\\ 
\textbf{Sociotech}   & 0.099                & 0.489               & 0.208                & -0.095 \\
$p$                  & (0.352)              & (\textbf{$<$0.0001})& (0.129)              & (0.452) \\
\textbf{Autonomy}    & ---                  & ---                 & ---                  & 0.138 \\
$p$                  &                      &                     &                      & (0.293) \\
\textbf{Feedback}    & ---                  & ---                 & ---                  & 0.593 \\
$p$                  &                      &                     &                      & (\textbf{$<$0.0001}) \\
\midrule
$R^2$                & 0.110                & 0.246               & 0.221                & 0.460 \\
Residual SE          & 0.804                & 0.815               & 1.047                & 0.875 \\
F-statistic          & 14.83$^{***}$        & 35.17$^{***}$       & 31.68$^{***}$        & 46.88$^{***}$ \\
\bottomrule
\end{tabular}
\label{tab:regression_results}
\end{table}

\subsection{Discriminant validity}
Discriminant validity has been defined as ``Two measures intended to measure distinct constructs have discriminant validity if the absolute value of the correlation between the measures after correcting for measurement error is low enough for the measures to be regarded as measuring distinct constructs''~\cite{ronkko2022updated}. To assess discriminant validity of our scale, we first computed the Heterotrait-Monotrait (HTMT) ratios of correlations~\cite{henseler2015new} between infrastructure scale dimensions and constructs used in criterion-related validity in the last subsection. The HTMT value between technical and sociotechnical infrastructure was 0.596. HTMT values between technical infrastructure and autonomy, feedback, and job satisfaction varied from 0.242 to 0.482. The same values for sociotechnical infrastructure were between 0.152 and 0.389. Thus, all HTMT values were well below the threshold of 0.85 advocated by Henseler~\cite{henseler2015new}, supporting discriminant validity among the constructs.

We compared the square roots of the average variance extracted (AVE) for technical and sociotechnical infrastructure ($\sqrt{\mathrm{AVE}}_{\mathit{tech}} = 0.793$; $\sqrt{\mathrm{AVE}}_{\mathit{sociotech}} = 0.787$) with the correlations from the model-implied latent correlation matrix of the CFA model, where the relationships between latent constructs are adjusted for measurement error. This is also called the Fornell–Larcker criterion~\cite{fornell1981evaluating}. The implied correlation from the CFA model between technical infrastructure and sociotechnical infrastructure was 0.631, while the correlations between other constructs were from 0.293 to 0.431. These results support discriminant validity of our scale, with both constructs sharing more variance with their own indicators than with any other construct.

Lastly, we followed the procedure advocated by R\"onkk\"o and Cho’s~\cite{ronkko2022updated} by comparing a CFA model in which the correlation between technical and sociotechnical infrastructure was freely estimated to a constrained model in which their correlation was fixed to 1.0. The chi-square difference test was significant, $\chi^2(1) = 170.1$, $p < .001$, and both AIC and BIC were lower for the unconstrained model (AIC: 2394.1 vs. 2562.1; BIC: 2478.6 vs. 2643.9). Taken together, all these results strongly indicate that the constructs of technical and sociotechnical infrastructure are empirically distinct and thus support discriminant validity.
\section{Discussion}\label{sec:discussion}

The presented Software Infrastructure Attitude Scale (SIAS) demonstrates strong psychometric validity, including construct, convergent, criterion-related, and discriminant validity. These properties were supported through expert review, item rating, and validation in a diverse sample of software professionals. Researchers have been interested in sociotechnical aspects of SE for over a decade~\cite{meneely2011socio}, with recent works focusing on these aspects further underlining their importance in SE research~\cite{hoda2021socio, lambiase2024cultural}. We believe our work provides a valuable tool for this line of research.

Both infrastructure dimensions performed as expected in relation to established constructs of autonomy, feedback from the job itself, and job satisfaction, with positive bivariate correlations with all. Attitudes toward technical infrastructure were associated with perceived autonomy, feedback from the job itself, and job satisfaction in multivariate regression analyses. These results are in line with self-determination theory (SDT)~\cite{deci2017self}, which emphasizes that employees who feel supported in their autonomy are more satisfied with their jobs. While autonomy did not predict job satisfaction in our model, this is likely due to shared variance with feedback. From the perspective of SDT, it would be logical that those SE professionals who have more positive attitudes toward technical infrastructure, have, on average, better technical infrastructure, which in turn facilitates higher autonomy, leading to higher job satisfaction.

Attitudes toward sociotechnical infrastructure only predicted feedback from the job itself, the definition of which is ``the degree to which carrying out the work activities required by the job results in the individual obtaining direct and clear information about the effectiveness of his or her performance''~\cite{hackman1976motivation}. It can be speculated that pieces of sociotechnical infrastructure such as coding standards and communication tools facilitate this feedback.

The practical contribution of this work is a scale with psychometric properties that researchers and practitioners can use to measure software professionals' attitudes toward technical and sociotechnical infrastructure. In empirical software engineering research, the scale can be included in survey studies, either as an independent or dependent variable, or as part of a model that combines it with a set of related constructs. This can mean that the scale can in part explain constructs such as job satisfaction, individual or team performance, well-being, or team dynamics. For example, we have already used the scale as a predictor of job satisfaction in a model explaining voluntary turnover~\cite{anon2025}. For industry, the scale can support efforts to evaluate and improve infrastructure-related aspects of the developer experience in organizational settings, for example, by comparing results across remote and on-site workers or by location.

The methodological contribution of this work is an exemplar paper in scale construction for a software engineering specific context. While prior scales for software engineering have been developed, they usually stop at an expert review or by just simply adapt scales from other contexts. We hope our work together with the guidelines by Graziotin et al.~\cite{graziotin2021psychometrics} inspires further work in software engineering scale development and improvement in measurement.

\subsection{Limitations}

The developed scale measures attitudes of software professionals toward technical and sociotechnical infrastructure, that is how it is perceived, not the quality of this infrastructure. Further empirical studies are needed to determine the relationship between perception and actual quality. Furthermore, we do not know much what professionals' attitude towards infrastructure affects, as one has to know how to measure the construct before it can be linked strongly to outcomes.

Overall attitude towards software infrastructure can be inferred by using the scale formatively~\cite{ralph2024teaching}, but attitude toward individual pieces of infrastructure cannot be inferred using this scale. Defining latent constructs is difficult, and our division into technical and sociotechnical infrastructure, while well justified, is not self-evident to practitioners. Thus, the definitions have to be shown to respondents when using the scale. Furthermore, the boundaries between these two constructs can be partially overlapping, as for example, version control systems can be used in technical and sociotechnical ways. Theorizing this distinction in software engineering would be valuable for future research.

One limitation of the current scale is the exclusion of all but one reverse-worded item. These items were originally included to control for response bias, but most of them exhibited low communalities and factor loadings. Thus, the resulting instrument includes mostly positively worded items, which may limit its robustness to inattentive responding. However, strong model fit and internal consistency of the refined scale mitigate this concern, and we further recommend using attention check items with the scale.

The scale was primarily constructed by academics, with additional items generated by a large language model (LLM). While the impact of using generative AI is unknown, all LLM-generated items underwent the same validation as human-generated ones. We found existing LLM guidelines~\cite{wagner2025towards} too exhaustive for our purposes. However, the usage is documented in the replication package. Of the 11 final items, 6 originated from the LLM (2 technical, 4 sociotechnical). Similarly, exploratory factor analysis involves subjective decisions, different cutoffs or their order could yield different results. We report our criteria: first removing cross-loading items, then those with low loadings. As expert evaluation is also subjective, we provide quantitative validation and document our expert review process in the replication package.

The affective content in the selected items, that is, information related to emotions, moods, and feelings, as discussed in Section~\ref{ssec:facevalidity}, is somewhat limited. Although some items do reflect affective components, they are fewer in numbers than items representing cognitive beliefs or past experiences. This may underrepresent the emotional dimension of attitudes in the resulting scale. This result emerged from following the scale development process, where items were retained based on psychometric properties rather than on theoretical balance. However, the tripartite theory of attitudes does not assume equal contributions from all components, and thus we view our scale as theoretically sound. 

Measurement invariance was not assessed in the current study. As our primary goal was to develop and validate the scale structure, future research should examine whether the factor structure and item loadings remain consistent across subgroups (e.g., by gender, region, or job role). Similarly, we did not assess test-retest reliability. It is financially difficult to assess measurement invariance before the scale is in use. Lastly, we have discussed some of the limitations in data gathering in a previous work~\cite{kuutila2025methodological}. 
\section{Conclusions}\label{sec:conclusions}
Software professionals are often interested in the technologies they use and the practices embedded in their teams’ ways of working, as evidenced by the popular Stack Overflow Developer Survey\footnote{https://survey.stackoverflow.co/2024/}. However, despite infrastructure being widely acknowledged as an important aspect of software development~\cite{storey2020software}, software engineering research has lacked validated tools to rigorously assess professionals’ attitudes toward it. Following the repeated calls for developing software engineering specific measures for behavioral and human factors research~\cite{gren2018standards, graziotin2021psychometrics}, we have developed two scales with psychometric properties to measure software professionals attitudes towards technical and sociotechnical infrastructure. Through rigorous item reduction and EFA, we identified a clear two-factor structure representing technical and sociotechnical dimensions, with strong loadings, low item complexity, and good model fit. CFA further supported factorial validity of the scale in a randomized subsample, showing excellent fit.

The content and face validity of the items were supported by their alignment with the cognitive, affective, and behavioral components of attitudes, and by expert judgment during item generation and refinement stages. Convergent validity was demonstrated through moderate to strong correlations with theoretically related constructs of autonomy, feedback from the job diagnostic survey and job characteristics model~\cite{hackman1976motivation, idaszak1987revision}, and job satisfaction. Criterion-related validity analyses showed that attitudes toward technical infrastructure significantly predicted job satisfaction, autonomy, and feedback. Finally, discriminant validity was established with HTMT ratios of correlations, the Fornell–Larcker criterion, and model comparison. 

These results provide strong empirical support for the reliability of the scale and validity of its structure. The scale offers a theoretically grounded and psychometrically supported tool for future research examining the role of workplace infrastructure in shaping employee attitudes and outcomes. Our work thus paves the way into theory-testing questionnaire research in the software engineering domain.

Future research can use this instrument to examine how attitudes toward workplace infrastructure shape a range of behaviors and outcomes. For instance, developers with positive attitudes toward technical and sociotechnical infrastructure may exhibit higher productivity, stronger organizational commitment, and more effective team coordination. They may also be more likely to engage in knowledge sharing, or produce higher-quality code. Our scale enables such hypotheses to be empirically tested for the first time in software engineering research.

%future work paragraph?

\section*{Data Availability}\label{sec:data_availability}
A replication package comprising (1) a dataset documenting the creation of the scale with item reduction through panel review and favourability ratings; (2) a dataset including data from participants who consented to their data being shared when administering the scale; (3) analysis scripts, is available at Zenodo~\cite{anonymous2025}.

\begin{acks}
This project was supported by Natural Sciences and Engineering Research Council of Canada Discovery Grant RGPIN-2020-05001, Discovery Accelerator Supplement RGPAS-2020-00081, and the Killam Foundation. We would like to thank Prof. Marcos Kalinowski, Prof. Nicole Novielli and Prof. Lucas Gren for their valuable contributions to this research.
\end{acks}

\bibliographystyle{ACM-Reference-Format}
\bibliography{sample-base}

@String{Computer = "{IEEE} Computer" }

@String{Psychometrika = "Psychometrika" }

@String{Academic = "Academic Press" }

@String{Springer = "Springer-Verlag" }

@article{graziotin2021psychometrics,
  title={Psychometrics in behavioral software engineering: A methodological introduction with guidelines},
  author={Graziotin, Daniel and Lenberg, Per and Feldt, Robert and Wagner, Stefan},
  journal={ACM Transactions on Software Engineering and Methodology (TOSEM)},
  volume={31},
  number={1},
  pages={1--36},
  doi={10.1145/3469888},
  year={2021},
  publisher={ACM New York, NY}
}

@book{trochim2001research,
  title={Research methods knowledge base},
  author={Trochim, William MK and Donnelly, James P},
  volume={2},
  year={2001},
  publisher={Atomic dog publishing Cincinnati, OH}
}

@article{storey2020software,
  title={The who, what, how of software engineering research: a socio-technical framework},
  author={Storey, Margaret-Anne and Ernst, Neil A and Williams, Courtney and Kalliamvakou, Eirini},
  journal={Empirical Software Engineering},
  volume={25},
  pages={4097--4129},
  year={2020},
  doi={10.1007/s10664-020-09858-z},
  publisher={Springer}
}

@article{hoda2021socio,
  title={Socio-technical grounded theory for software engineering},
  author={Hoda, Rashina},
  journal={IEEE Transactions on Software Engineering},
  volume={48},
  number={10},
  pages={3808--3832},
  doi={10.1109/TSE.2021.3106280},
  year={2021},
  publisher={IEEE}
}

@article{judge2000personality,
  title={Personality and job satisfaction: the mediating role of job characteristics.},
  author={Judge, Timothy A and Bono, Joyce E and Locke, Edwin A},
  journal={Journal of applied psychology},
  volume={85},
  number={2},
  pages={237},
  year={2000},
  doi={10.1037/0021-9010.85.2.237},
  publisher={American Psychological Association}
}

@article{rosseel2012lavaan,
  title={lavaan: An R package for structural equation modeling},
  author={Rosseel, Yves},
  journal={Journal of statistical software},
  volume={48},
  pages={1--36},
  year={2012}
}

@incollection{ralph2024teaching,
  title={Teaching Software Metrology: The Science of Measurement for Software Engineering},
  author={Ralph, Paul and Kuutila, Miikka and Arif, Hera and Ayoola, Bimpe},
  booktitle={Handbook on Teaching Empirical Software Engineering},
  pages={101--154},
  doi={10.1007/978-3-031-71769-7_5},
  year={2024},
  publisher={Springer}
}

@article{fabrigar1999evaluating,
  title={Evaluating the use of exploratory factor analysis in psychological research.},
  author={Fabrigar, Leandre R and Wegener, Duane T and MacCallum, Robert C and Strahan, Erin J},
  journal={Psychological methods},
  volume={4},
  number={3},
  pages={272},
  year={1999},
  publisher={American Psychological Association}
}

@book{haddock2004contemporary,
  title={Contemporary perspectives on the psychology of attitudes},
  author={Haddock, Geoffrey and Maio, Gregory R and others},
  year={2004},
  publisher={Psychology Press Hove}
}

@article{russell2003core,
  title={Core affect and the psychological construction of emotion.},
  author={Russell, James A},
  journal={Psychological review},
  volume={110},
  number={1},
  pages={145},
  year={2003},
  publisher={American Psychological Association}
}

@book{fishbein1977belief,
  title={Belief, attitude, intention, and behavior: An introduction to theory and research},
  author={Fishbein, Martin and Ajzen, Icek},
  publisher={Reading, MA:Addison-Wesley},
  year={1977}
}

@Inbook{ajzen1985intentions,
author="Ajzen, Icek",
editor="Kuhl, Julius
and Beckmann, J{\"u}rgen",
title="From Intentions to Actions: A Theory of Planned Behavior",
bookTitle="Action Control: From Cognition to Behavior",
year="1985",
publisher="Springer Berlin Heidelberg",
address="Berlin, Heidelberg",
pages="11--39",
isbn="978-3-642-69746-3",
doi="10.1007/978-3-642-69746-3_2",
url="https://doi.org/10.1007/978-3-642-69746-3_2"
}

@article{davis1989perceived,
  title={Perceived usefulness, perceived ease of use, and user acceptance of information technology},
  author={Davis, Fred D},
  journal={MIS quarterly},
  pages={319--340},
  year={1989},
  doi={10.2307/249008},
  publisher={JSTOR}
}

@article{venkatesh2003user,
  title={User acceptance of information technology: Toward a unified view},
  author={Venkatesh, Viswanath and Morris, Michael G and Davis, Gordon B and Davis, Fred D},
  journal={MIS quarterly},
  pages={425--478},
  year={2003},
  doi={10.2307/30036540},
  publisher={JSTOR}
}

@book{heider2013psychology,
  title={The psychology of interpersonal relations},
  author={Heider, Fritz},
  year={2013},
  publisher={Psychology Press}
}

@article{kitchen2014elaboration,
  title={The elaboration likelihood model: review, critique and research agenda},
  author={Kitchen, Philip J and Kerr, Gayle and Schultz, Don E and McColl, Rod and Pals, Heather},
  journal={European Journal of marketing},
  volume={48},
  number={11/12},
  pages={2033--2050},
  year={2014},
  doi={10.1108/EJM-12-2011-0776},
  publisher={Emerald Group Publishing Limited}
}

@book{blumer1986symbolic,
  title={Symbolic interactionism: Perspective and method},
  author={Blumer, Herbert},
  year={1986},
  publisher={Univ of California Press}
}

@article{katz1960functional,
  title={The functional approach to the study of attitudes},
  author={Katz, Daniel},
  journal={Public opinion quarterly},
  volume={24},
  number={2},
  pages={163--204},
  year={1960},
  doi={10.1086/266945},
  publisher={Oxford University Press}
}

@book{festinger1957theory,
  title={A theory of cognitive dissonance},
publisher={Stanford university press},
  author={Festinger, Leon},
  year={1957}
}

@book{harmon2019introduction,
  title={An introduction to cognitive dissonance theory and an overview of current perspectives on the theory.},
  author={Harmon-Jones, Eddie and Mills, Judson},
  year={2019},
  publisher={American Psychological Association}
}

@article{atkinson1957motivational,
  title={Motivational determinants of risk-taking behavior.},
  author={Atkinson, John W},
  journal={Psychological review},
  volume={64},
  number={6p1},
  pages={359},
  year={1957},
  doi={10.1037/h0043445},
  publisher={American Psychological Association}
}

@article{stern1999value,
  title={A value-belief-norm theory of support for social movements: The case of environmentalism},
  author={Stern, Paul C and Dietz, Thomas and Abel, Troy and Guagnano, Gregory A and Kalof, Linda},
  journal={Human ecology review},
  pages={81--97},
  year={1999},
  publisher={JSTOR}
}

@book{bandura1986social,
  title={Social foundations of thought and action},
  author={Bandura, Albert},
  publisher={Prentice Hall},
  year={1986}
}

@incollection{eccles1983expectancies,
  title={Expectancies, values, and academic behaviors},
  author={Eccles, Jacquelynne S},
  booktitle={Achievement and achievement motives},
  pages={75--146},
  year={1983},
  publisher={Freeman}
}

@article{idaszak1987revision,
  title={A revision of the Job Diagnostic Survey: Elimination of a measurement artifact.},
  author={Idaszak, Jacqueline R and Drasgow, Fritz},
  journal={Journal of applied psychology},
  volume={72},
  number={1},
  pages={69},
  year={1987},
  publisher={American Psychological Association}
}

@incollection{petty1986elaboration,
  title={The elaboration likelihood model of persuasion},
  author={Petty, Richard E and Cacioppo, John T},
  booktitle={Advances in experimental social psychology},
  volume={19},
  pages={123--205},
  year={1986},
  doi={10.1016/S0065-2601(08)60214-2},
  publisher={Elsevier}
}

@article{worthington2006scale,
  title={Scale development research: A content analysis and recommendations for best practices},
  author={Worthington, Roger L and Whittaker, Tiffany A},
  journal={The counseling psychologist},
  volume={34},
  number={6},
  pages={806--838},
  year={2006},
  doi={10.1177/0011000006288127},
  publisher={Sage Publications Sage CA: Thousand Oaks, CA}
}

@book{devellis2021scale,
  title={Scale development: Theory and applications},
  author={DeVellis, Robert F and Thorpe, Carolyn T},
  year={2021},
  publisher={Sage publications}
}

@article{graziotin2014happy,
  title={Happy software developers solve problems better: psychological measurements in empirical software engineering},
  author={Graziotin, Daniel and Wang, Xiaofeng and Abrahamsson, Pekka},
  journal={PeerJ},
  volume={2},
  pages={e289},
  doi={10.7717/peerj.289},
  year={2014},
  publisher={PeerJ Inc.}
}

@inproceedings{gren2018standards,
  title={Standards of validity and the validity of standards in behavioral software engineering research: the perspective of psychological test theory},
  author={Gren, Lucas},
  booktitle={Proceedings of the 12th ACM/IEEE international symposium on empirical software engineering and measurement},
  pages={1--4},
  year={2018}
}

@article{revelle2015package,
  title={Package ‘psych’},
  author={Revelle, William},
  journal={The comprehensive R archive network},
  volume={337},
  number={338},
  pages={161--165},
  year={2015}
}

@article{hackman1976motivation,
  title={Motivation through the design of work: Test of a theory},
  author={Hackman, J Richard and Oldham, Greg R},
  journal={Organizational behavior and human performance},
  volume={16},
  number={2},
  pages={250--279},
  year={1976},
  doi={10.1016/0030-5073(76)90016-7},
  publisher={Elsevier}
}

@article{cronbach1955construct,
  title={Construct validity in psychological tests.},
  author={Cronbach, Lee J and Meehl, Paul E},
  journal={Psychological bulletin},
  volume={52},
  number={4},
  pages={281},
  year={1955},
  publisher={American Psychological Association}
}

@book{apa_dictionary,
  title     = {APA Dictionary of Psychology},
  publisher = {American Psychological Association},
  year      = {2025},
  edition   = {2nd},
  url   = {https://dictionary.apa.org/construct-validity},
  note = {Entry: [construct validity]}
}

@article{OSTROM196912,
title = {The relationship between the affective, behavioral, and cognitive components of attitude},
journal = {Journal of Experimental Social Psychology},
volume = {5},
number = {1},
pages = {12-30},
year = {1969},
issn = {0022-1031},
doi = {https://doi.org/10.1016/0022-1031(69)90003-1},
url = {https://www.sciencedirect.com/science/article/pii/0022103169900031},
author = {Thomas M. Ostrom}
}

@article{hanseth2001designing,
  title={Designing work oriented infrastructures},
  author={Hanseth, Ole and Lundberg, Nina},
  journal={Computer Supported Cooperative Work (CSCW)},
  volume={10},
  pages={347--372},
  doi={10.1023/A:1012727708439},
  year={2001},
  publisher={Springer}
}

@article{ottens2006modelling,
  title={Modelling infrastructures as socio-technical systems},
  author={Ottens, Maarten and Franssen, Maarten and Kroes, Peter and Van De Poel, Ibo},
  journal={International journal of critical infrastructures},
  volume={2},
  number={2-3},
  pages={133--145},
  doi={10.1504/IJCIS.2006.009433},
  year={2006},
  publisher={Inderscience Publishers}
}

@book{winters2020software,
  title={Software engineering at google: Lessons learned from programming over time},
  author={Winters, Titus and Manshreck, Tom and Wright, Hyrum},
  year={2020},
  publisher={O'Reilly Media, Inc.}
}

@article{westfall2016statistically,
  title={Statistically controlling for confounding constructs is harder than you think},
  author={Westfall, Jacob and Yarkoni, Tal},
  journal={PloS one},
  volume={11},
  number={3},
  pages={e0152719},
  year={2016},
  doi={10.1371/journal.pone.0152719},
  publisher={Public Library of Science San Francisco, CA USA}
}

@article{nisbett1977halo,
  title={The halo effect: Evidence for unconscious alteration of judgments.},
  author={Nisbett, Richard E and Wilson, Timothy D},
  journal={Journal of personality and social psychology},
  volume={35},
  number={4},
  pages={250},
  year={1977},
  doi={10.1037/0022-3514.35.4.250},
  publisher={American Psychological Association}
}

@article{lyu2025systematic,
  title={A Systematic Literature Review of Infrastructure Studies in SIGCHI},
  author={Lyu, Yao and Cai, Jie and Carroll, John M},
  journal={arXiv preprint arXiv:2504.09612},
  year={2025}
}

@article{agenor2010theory,
  title={A theory of infrastructure-led development},
  author={Ag{\'e}nor, Pierre-Richard},
  journal={Journal of Economic Dynamics and Control},
  volume={34},
  number={5},
  pages={932--950},
  year={2010},
  doi={j.jedc.2010.01.009},
  publisher={Elsevier}
}

@article{dutra2022tact,
  title={TACT: An insTrument to Assess the organizational ClimaTe of agile teams-A Preliminary Study},
  author={Dutra, Eliezer and Lima, Patr{\'\i}cia and Cerdeiral, Cristina and Diirr, Bruna and Santos, Gleison},
  journal={Journal of Software Engineering Research and Development},
  volume={10},
  doi={10.5753/jserd.2021.1973},
  pages={1--1},
  year={2022}
}

@inproceedings{mattos2024instrument,
  title={An Instrument for Assessing Power Distance in Agile Organizations-Preliminary Results},
  author={Mattos, Claudio Saraiva and Dutra, Eliezer and Santos, Gleison},
  booktitle={Proceedings of the XXIII Brazilian Symposium on Software Quality},
  pages={136--146},
  doi={10.1145/3701625.3701653},
  year={2024}
}

@article{wilkinson2019towards,
  title={Towards an archaeological theory of infrastructure},
  author={Wilkinson, Darryl},
  journal={Journal of Archaeological Method and Theory},
  volume={26},
  number={3},
  pages={1216--1241},
  year={2019},
  doi={10.1007/s10816-018-9410-2},
  publisher={Springer}
}

@inproceedings{star1996steps,
  title={Steps towards an ecology of infrastructure: complex problems in design and access for large-scale collaborative systems},
  author={Star, Susan Leigh and Ruhleder, Karen},
  booktitle={Proceedings of the 1994 ACM Conference on Computer Supported Cooperative Work},
  pages={253--256},
  year={1994},
  doi={10.1145/192844.193021},
  numpages = {12},
  series = {CSCW '94}
}

@article{star1999ethnography,
  title={The ethnography of infrastructure},
  author={Star, Susan Leigh},
  journal={American behavioral scientist},
  volume={43},
  number={3},
  pages={377--391},
  doi={10.1177/00027649921955326},
  year={1999},
  publisher={Sage Publications, Inc.}
}

@techreport{buhr2003infrastructure,
  title={What is infrastructure?},
  author={Buhr, Walter},
  year={2003},
  institution={Volkswirtschaftliche Diskussionsbeitr{\"a}ge}
}

@article{graziotin2015feelings,
  title={Do feelings matter? On the correlation of affects and the self-assessed productivity in software engineering},
  author={Graziotin, Daniel and Wang, Xiaofeng and Abrahamsson, Pekka},
  journal={Journal of Software: Evolution and Process},
  volume={27},
  number={7},
  pages={467--487},
  doi={10.1002/smr.1673},
  year={2015},
  publisher={Wiley Online Library}
}

@article{novielli2019sentiment,
  title={Sentiment and emotion in software engineering},
  author={Novielli, Nicole and Serebrenik, Alexander},
  journal={IEEE Software},
  volume={36},
  number={5},
  pages={6--23},
  doi={10.1109/MS.2019.2924013},
  year={2019},
  publisher={IEEE}
}

@article{williams2010exploratory,
  title={Exploratory factor analysis: A five-step guide for novices},
  author={Williams, Brett and Onsman, Andrys and Brown, Ted},
  journal={Australasian journal of paramedicine},
  volume={8},
  pages={1--13},
  year={2010},
  doi={10.33151/ajp.8.3.93},
  publisher={SAGE Publications Sage UK: London, England}
}

@article{kaiser1974index,
  title={An index of factorial simplicity},
  author={Kaiser, Henry F},
  journal={Psychometrika},
  volume={39},
  number={1},
  pages={31--36},
  doi={10.1007/BF02291575},
  year={1974},
  publisher={Springer}
}

@article{weijters2012misresponse,
  title={Misresponse to reversed and negated items in surveys: A review},
  author={Weijters, Bert and Baumgartner, Hans},
  journal={Journal of Marketing Research},
  volume={49},
  number={5},
  pages={737--747},
  year={2012},
  doi={10.1509/jmr.11.0368},
  publisher={SAGE Publications Sage CA: Los Angeles, CA}
}

@article{henseler2015new,
  title={A new criterion for assessing discriminant validity in variance-based structural equation modeling},
  author={Henseler, J{\"o}rg and Ringle, Christian M and Sarstedt, Marko},
  journal={Journal of the academy of marketing science},
  volume={43},
  pages={115--135},
  doi={10.1007/s11747-014-0403-8},
  year={2015},
  publisher={Springer}
}

@article{ronkko2022updated,
  title={An updated guideline for assessing discriminant validity},
  author={R{\"o}nkk{\"o}, Mikko and Cho, Eunseong},
  journal={Organizational research methods},
  volume={25},
  number={1},
  pages={6--14},
  year={2022},
  doi={10.1177/1094428120968614},
  publisher={Sage Publications Sage CA: Los Angeles, CA}
}

@article{fornell1981evaluating,
  title={Evaluating structural equation models with unobservable variables and measurement error},
  author={Fornell, Claes and Larcker, David F},
  journal={Journal of marketing research},
  volume={18},
  number={1},
  pages={39--50},
  year={1981},
  publisher={Sage Publications Sage CA: Los Angeles, CA}
}

@article{vrieze2012model,
  title={Model selection and psychological theory: a discussion of the differences between the Akaike information criterion (AIC) and the Bayesian information criterion (BIC).},
  author={Vrieze, Scott I},
  journal={Psychological methods},
  volume={17},
  number={2},
  pages={228},
  year={2012},
  doi={10.1037/a0027127},
  publisher={American Psychological Association}
}

@article{baltes2022sampling,
  title={Sampling in software engineering research: A critical review and guidelines},
  author={Baltes, Sebastian and Ralph, Paul},
  journal={Empirical Software Engineering},
  volume={27},
  number={4},
  pages={94},
  year={2022},
  doi={10.1007/s10664-021-10072-8},
  publisher={Springer}
}

@book{oppenheim2000questionnaire,
  title={Questionnaire design, interviewing and attitude measurement},
  author={Oppenheim, Abraham Naftali},
  year={2000},
  publisher={Bloomsbury Publishing}
}

@article{dolnicar2014including,
  title={Including Don't know answer options in brand image surveys improves data quality},
  author={Dolnicar, Sara and Gr{\"u}n, Bettina},
  journal={International Journal of Market Research},
  volume={56},
  number={1},
  pages={33--50},
  year={2014},
  doi={10.2501/IJMR-2013-043},
  publisher={SAGE Publications Sage UK: London, England}
}

@article{kung2018attention,
  title={Are attention check questions a threat to scale validity?},
  author={Kung, Franki YH and Kwok, Navio and Brown, Douglas J},
  journal={Applied Psychology},
  volume={67},
  number={2},
  pages={264--283},
  year={2018},
  doi={10.1111/apps.12108},
  publisher={Wiley Online Library}
}

@article{malhotra2008completion,
  title={Completion time and response order effects in web surveys},
  author={Malhotra, Neil},
  journal={Public opinion quarterly},
  volume={72},
  number={5},
  pages={914--934},
  year={2008},
  doi={10.1093/poq/nfn050},
  publisher={Oxford University Press}
}

@article{DBLP:journals/software/DAngeloLDEHGJ24,
  author       = {Sarah D'Angelo and
                  Jessica Lin and
                  Jill Dicker and
                  Carolyn D. Egelman and
                  Maggie Hodges and
                  Collin Green and
                  Ciera Jaspan},
  title        = {Measuring Developer Experience With a Longitudinal Survey},
  journal      = {{IEEE} Softw.},
  volume       = {41},
  number       = {4},
  pages        = {19--24},
  year         = {2024},
  url          = {https://doi.org/10.1109/MS.2024.3386027},
  doi          = {10.1109/MS.2024.3386027},
  bibsource    = {dblp computer science bibliography, https://dblp.org}
}

@inproceedings{anon2025,
  author    = {Kuutila, Miikka and Ralph, Paul and Qiu, Huilian Sophie and de Souza Santos, Ronnie and Choetkiertikul, Morakot and Fard, Amin Milani and Alkadhi, Rana and Devroey, Xavier and Robles, Gregorio and Hata, Hideaki and Baltes, Sebastian and Arif, Hera and Kovalenko, Vladimir and Chakraborty, Shalini and Tuzun, Eray and Adisaputri, Gianisa},
  title     = {Staying or Leaving? How Job Satisfaction, Embeddedness and Antecedents Predict Turnover Intentions of Software Professionals},
  booktitle = {Proceedings of the 2026 IEEE/ACM 48th International Conference on Software Engineering (ICSE '26)},
  pages = {1--13},
  location = {Rio de Janeiro, Brazil},
  publisher = {ACM},
  address = {New York, NY, USA},
  year      = {2026},
  doi = {10.1145/3744916.3773124}
}

@article{lenberg2015behavioral,
  title={Behavioral software engineering: A definition and systematic literature review},
  author={Lenberg, Per and Feldt, Robert and Wallgren, Lars G{\"o}ran},
  journal={Journal of Systems and software},
  volume={107},
  pages={15--37},
  year={2015},
  doi={10.1016/j.jss.2015.04.084},
  publisher={Elsevier}
}

@dataset{anonymous2025,
  author       = {Kuutila, Miikka and Ralph, Paul and Qiu, Huilian Sophie and de Souza Santos, Ronnie and Choetkiertikul, Morakot and Milani Fard, Amin and Alkadhi, Rana and Devroey, Xavier and Robles, Gregorio and Hata, Hideaki and Baltes, Sebastian and Kovalenko, Vladimir and Chakraborty, Shalini and Tuzun, Eray and Arif, Hera and Adisaputri, Gianisa and Garcés, Kelly and Severo Lisbôa de Andrade, Anielle and Amedzor, Eyram and Ayoola, Bimpe and Gaspard-Chickoree, Keisha and Hoseyni, Arazoo},
  title        = {Replication package for the paper "The Software Infrastructure Attitude Scale (SIAS): A Questionnaire Instrument for Measuring Professionals' Attitudes Toward Technical and Sociotechnical Infrastructure"},
  month        = jul,
  year         = 2025,
  publisher    = {Zenodo},
  doi          = {10.5281/zenodo.15962431},
  url          = {https://doi.org/10.5281/zenodo.15962431},
}

@article{jebb2021review,
  title={A review of key Likert scale development advances: 1995--2019},
  author={Jebb, Andrew T and Ng, Vincent and Tay, Louis},
  journal={Frontiers in psychology},
  volume={12},
  pages={637547},
  doi={10.3389/fpsyg.2021.637547},
  year={2021},
  publisher={Frontiers Media SA}
}

@article{deci2017self,
  title={Self-determination theory in work organizations: The state of a science},
  author={Deci, Edward L and Olafsen, Anja H and Ryan, Richard M},
  journal={Annual review of organizational psychology and organizational behavior},
  volume={4},
  pages={19--43},
  year={2017},
  doi={10.1146/annurev-orgpsych-032516-113108},
  publisher={Annual Review of Organizational Psychology and Organizational Behavior}
}

@INPROCEEDINGS{wagner2025towards,
  author={Wagner, Stefan and Barón, Marvin Muñoz and Falessi, Davide and Baltes, Sebastian},
  booktitle={2025 IEEE/ACM International Workshop on Methodological Issues with Empirical Studies in Software Engineering (WSESE)}, 
  title={Towards Evaluation Guidelines for Empirical Studies Involving LLMs}, 
  year={2025},
  volume={},
  number={},
  pages={24-27},
  doi={10.1109/WSESE66602.2025.00011}}

@article{enders2001relative,
  title={The relative performance of full information maximum likelihood estimation for missing data in structural equation models},
  author={Enders, Craig K and Bandalos, Deborah L},
  journal={Structural Equation Modeling: A Multidisciplinary Journal},
  volume={8},
  number={3},
  pages={430--457},
  doi={10.1207/S15328007SEM0803\_5},
  year={2001},
  publisher={Taylor \& Francis}
}

@article{haynes1995content,
  title={Content validity in psychological assessment: A functional approach to concepts and methods.},
  author={Haynes, Stephen N and Richard, David and Kubany, Edward S},
  journal={Psychological assessment},
  volume={7},
  number={3},
  pages={238},
  year={1995},
  doi={10.1037/1040-3590.7.3.238},
  publisher={American Psychological Association}
}

@article{choi2004catalog,
  title={A catalog of biases in questionnaires},
  author={Choi, Bernard CK and Pak, Anita WP},
  journal={Preventing chronic disease},
  volume={2},
  number={1},
  pages={A13},
  year={2004}
}

@inproceedings{kuutila2025methodological,
  title={Methodological and Practical Challenges in Longitudinal, Large-Scale, Collaborative Questionnaire Survey Research},
  author={Kuutila, Miikka and Ralph, Paul},
  booktitle={2025 IEEE/ACM International Workshop on Methodological Issues with Empirical Studies in Software Engineering (WSESE)},
  pages={64--69},
  doi={10.1109/WSESE66602.2025.00017},
  year={2025},
  organization={IEEE}
}

@inproceedings{lambiase2024cultural,
  title={Cultural and socio-technical aspects in software development},
  author={Lambiase, Stefano},
  booktitle={Proceedings of the 28th International Conference on Evaluation and Assessment in Software Engineering},
  pages={482--487},
  doi={10.1145/3661167.3661230},
  year={2024}
}

@inproceedings{meneely2011socio,
  title={Socio-technical developer networks: Should we trust our measurements?},
  author={Meneely, Andrew and Williams, Laurie},
  booktitle={Proceedings of the 33rd International Conference on Software Engineering},
  pages={281--290},
  doi={10.1145/1985793.1985832},
  year={2011}
}

\end{document}